\documentclass[12pt,preprint]{aastex}

\begin{document}

\title{
S\'{e}rsic 159-03: discovery of the brightest soft X--ray excess emitting  cluster of galaxies
}
\author{Massimiliano Bonamente$\,^{1}$,Richard~Lieu$\,^{1}$ and
Jonathan~P.~D.~Mittaz$\,^{2}$}

\affil{\(^{\scriptstyle 1} \){Department of Physics, University of Alabama,
Huntsville, AL 35899, U.S.A.}\\
\(^{\scriptstyle 2} \){Mullard Space Science Laboratory, UCL,
Holmbury St. Mary, Dorking, Surrey, RH5 6NT, U.K.}\\
}

\begin{abstract}
The soft X-ray excess emission in the southern cluster S\'{e}rsic 159-03
represents hiterto the strongest effect of its kind.
Emission in the $\sim 0.2-0.4$ keV passband is detected 
far in excess of the expected contribution from
the hot phase of the intra-cluster medium, and 
extends to the X-ray signal limit of the 
cluster.
Our analysis of {\it ROSAT} PSPC observations reveal that the {\it soft excess} can be
interpreted as either a thermal or non-thermal effect, and
the high data quality allows to place tight constraints on the 
two currently competing models.
However, 
each model now
implies major revisions to our understanding of clusters of galaxies: either
`warm' gas masses in similar amounts to the hot gas, or relativistic particles
in or above equipartition with the hot phase, appear to be 
unavoidable.
\end{abstract}


\section*{Introduction}
During the five intervening years since the discovery of EUV and
soft X--ray excess 
emission in the Virgo and Coma clusters with the {\it Extreme Ultraviolet
Explorer} (EUVE) and {\it ROSAT} missions (Lieu et al. 1996a,b), the original results have been
cross-checked by follow-up observations of the same objects plus a larger sample of
nearby galaxy clusters (Fabian 1996; Bonamente, Lieu and Mittaz. 2001a,b; Bonamente et al. 2001c; 
Mittaz, Lieu and Lockman 1998; Lieu 
et al. 1999; Lieu, Bonamente and Mittaz 1999,2000; Bergh\"{o}fer, Bowyer and Korpela 2000a,b;
Bowyer and Bergh\"{o}fer 1998; Bowyer, Korpela and Bergh\"{o}fer 2001;  Bowyer,  Bergh\"{o}fer
and Korpela 1999; Arabadjis and Bregman 1999; Reynolds et al. 1999; Kaastra et al. 1999;
Dixon, Sallmen and Hurwitz 2001; Dixon, Hurwitz and Ferguson 1996;
Valinia et al. 2000; Fusco-Femiano et al. 2000
 Buote 2000a,b,2001; Arnaud et al. 2001), using complementary data from the
{\it BeppoSAX}, {\it HUT}, {\it FUSE} and {\it RXTE} missions.
Recently, we have undertaken a more extensive study of soft X-ray emission in galaxy clusters
of low/intermediate $N_H$, with the aim of assessing the impact and cosmological significance 
of the {\it soft excess} phenomenon. Our efforts led to the discovery of soft excess emission
in a sample of clusters in the Shapley concentration (Bonamente et al. 2001b), and while 
the analysis of more sources is still underway, we report in this {\it Letter} the 
brightest soft X-ray ($\sim 0.2-0.4$ keV) excess cluster within the {\it ROSAT} PSPC database. 
The excess emission reaches $\sim$ 100 \% of the hot intra--cluster medium 
(ICM) contribution, 3-4 times higher than the same for any other PSPC  targets
(e.g., A1795 and Virgo; Bonamente, Lieu and Mittaz 2001a).  
The inverse-Compton (IC) interpretation of the soft emission (see section 3) results
in a non-thermal pressure that greatly exceeds that of the thermal gas: the excess of S\'{e}rsic 159-03
has more demanding energetic requirements than all previously known cases (e.g.,
Coma, Virgo and A1795; Lieu et al. 1999, Bonamente, Lieu and Mittaz 2001a). 
Alternatively, if the origin is thermal,
the mass implications of the putative `warm' gas are also
extreme, with the warm-to-hot gas mass ratio $\sim$ 50 \% (this figure is
matched only by the emission seen by EUVE for A2199, however at lower statistical significance;
Lieu, Bonamente and Mittaz 2000).

\section*{X-ray, infrared and 21 centimeter observations of S\'{e}rsic 159-03}
S\'{e}rsic 159-03 is a rich southern galaxy cluster, 
also known as Abell S1101,
at a redshift of z=0.056 (Abell, Corwin and Olowin 1989). 
The cluster's X-ray luminosity (L$_X = 5.35 \times 10^{44}$ erg/s in 0.5-2.0 keV,
De Grandi et al. 1999) is typical for a cluster of its temperature (e.g., Wu, Xue and Fang 1999),
and the emission has a radial extent of $\sim$ 10 arcmin (Fig. 1). Recent XMM observations
(Kaastra et al. 2001) show a `cooling' of the hot ICM in the central regions and
no obvious peculiarities in the X-ray morphology, consistent with our PSPC data.
The cluster, moreover, has no known radio emissions associated with it.

By means of three pointed PSPC observations of this source, taken between May 1992 and 
May 1993 with a total
exposure of about 19,000 seconds, we found a very bright excess emission
in PSPC's 1/4 keV band (also known as R2 band, $\sim 0.2-0.4$ keV). Fig. 2 shows how
detected emission (light blue, dark blue and black crosses) compare with the
expected emission from the hot intra--cluster medium (ICM, solid lines), e.g. as described
by the recent X--ray analysis of  Kaastra et al. (2001).
The three observations were modeled simultaneously, in order 
to reduce statistical uncertainties. 

As extragalactic soft X--ray emissions are absorbed by the interstellar medium,
an accurate measurement of the line-of-sight 
Galactic hydrogen (HI) column density ($N_H$) is crucial. The Dickey and Lockman (1990) HI maps
at resolutions of respectively 1, 4 and 9 square degrees centered at the cluster's position 
indicate a region of sky free from spatial HI gradients, with
an HI density at the cluster position of 1.79 $\times 10^{20}$ cm$^{-2}$.
Since Galactic infrared 100 $\mu$m emission correlates well with HI 21-cm radiation 
(Boulanger and Perault 1988),
we further consulted {\it IRAS} 100 $\mu$m maps (Wheelock et al. 1994)
at a finer resolution of 3 arcmin.  
The 100 $\mu$m emission is very smooth over the cluster's region,
and it confirmed our $N_H$ value for the cluster's Galactic HI.
This value is henceforth used in the spectral analysis~\footnote{The low declination ($\sim -42^o$)
renders this source not easily  observable with the NRAO 43-mt Green Bank telescope we employed
in our previous analyses.}. We refer to a previous paper (Bonamente et al. 2001b)
for the details on PSPC data analysis techniques.

Spectra of azimuthally averaged annular regions were modelled with a single temperature
optically-thin plasma emission.~\footnote{The 
spectral analysis was performed with the XSPEC software. 
PSPC pulse-invariant (PI) channels
1-19 and 202-256 are not used, as potentially affected by calibration uncertainties.}
The Galactic absorption was modelled with the codes of Morrison and MacCammon 1983 (MM83; `WABS' in XSPEC
language) and of Wilms, Allen and McCray 2000 (WAM00; `TBABS' in XSPEC). The He cross-sections
of MM83 are in good agreement with the recent compilation by Yan, Sadeghpour and Dalgarno (1998), as
also shown in a previous paper (Bonamente et al. 2001a). The WAM00 code provides 
however the most up-to-date
cross-section compilation for absorption of X-rays in the interstellar medium.
 In the energy
range of interest, 0.2-0.4 keV, the two codes are nearly indistinguishable at the
resolution of the PSPC (see Fig. 4 of 
WAM00), and yielded statistically consistent results, see Tables 1 and 2. The emission
code employed was MEKAL (after Mewe, Gronenschild and van den Oord, 1985;
 Mewe, Lemen and van den Oord 1986; Kaastra 1992). 

When applied to the entire passband of  the PSPC (0.2-2.0 keV), 
the fit is formally unacceptable 
(e.g., red crosses in Fig. 3), and a region of depleted
flux (0.5-1.0 keV) appears to accompany the 0.2-0.4 keV excess. The fit is satisfactory
when applyed only to energies $\geq$ 0.5 keV, with the soft excess
then abundantly  revealed at E$\leq$ 0.4 keV (green crosses, Fig. 3). The best-fit parameters
of the hot phase agree with the most recent X--ray 
{\it XMM} observation (Kaastra et al. 2001). 
In Table 1 we 
summarise the results.
We emphasize that the soft excess emission is so strong that, in order to be interpreted as
the low--energy tail of the hot ICM emission, a Galactic column density as low as
5-8 $\times 10^{19}$ cm$^{-2}$ would be required, which is  not only severely discrepant
from the
measured
values, but even lower that that of the `Lockman hole' where $N_H$ reaches global minimum.
The soft emission cannot therefore be attributed to peculiarities in the Galactic 
HI distribution.

In Figure 4 we show the trend of the
PSPC excess. The 
 emission extends to a radius of 800 kiloparsec~\footnote{A Hubble constant
of $H_0=50$ km s$^{-1}$ Mpc$^{-1}$ is assumed here and after.} (=9 arcmin); in the 9--12 arcmin
region, where only marginal cluster emission is detected (Fig. 1), the data point
is consistent with zero flux, thereby confirming the integrity of our
background subtraction technique (Bonamente et al. 2001b for reference).
As comparison, we note that from  the {\it XMM}/EPIC data
of Kaastra et al. (2001)  evidence for X-ray emission
is apparent out to a radial distance of about 10 arcmin, in agreement
with our PSPC data. The soft excess
emission therefore persists to the X-ray detection limit. 
To illustrate the significance of the role played by the model
uncertanties in our determination of the 1/4 keV contribution from the hot ICM,
we consider as example the 3-6' region. When the statistical error in the 
0.5-2.0 keV PSPC data (the portion of the spectra used
for model determination) is taken into account, the error in the  fractional eccess
increases only 
from 9 \% (see Fig. 4) to 10\%.
Similar conclusions apply to the other regions, indicating that 
evidence for a soft X-ray component from
S\'{e}rsic 159-03 is unaffected by statistical uncertanties in the hot ICM modelling 
~\footnote{Krick, Arabadjis and Bregman (2000) 
argue that sub-arcmin variations of $N_H$ could in
principle account for a 20-30 \% excess. Although the relevance of  this phenomenon
to our S\'{e}rsic 159-03 observations is uncertain, the strength of the PSPC excess here reported
largely exceeds that figure. On occasions where comparison with stellar Lyman-$\alpha$ 
and quasar X-ray spectra could be made, the agreement among the different methods of
determining $N_H$ would suggest an error less than $\sim 10^{19}$ cm$^{-2}$
(see Lieu et al. 1996b and references therein).}.

\section*{Interpretation}

How do we interpret such a  strong signal?
A `warm', optically--thin phase of the intra-cluster medium (ICM) 
at sub-megakelvin temperatures can in principle accont for the
excess emission.
Thermal conduction can be suppressed, either for the presence of tangled magnetic fields
(Chandran et al. 1999) or if the warm phase is sufficiently `clumpy' 
(Cowie and McKee 1977; Smith and Lilliequist 1979),
yet the time scale for radiative losses of  a 
`warm' gas
is sufficiently short to  necessitate  rapid and continuous replenishment
(such as that of the `mixing layers' scenario of Fabian 1997), if the
warm phase existed in a steady state for cosmological times.
In the neighborhood of  T $\sim 10^6$ K, the cooling time of a gas with solar abundances
is $t_{cool} \sim 3 \times 10^7 \times n_{-3}^{-1}$ years, where $n_{-3}$
is gas density in units of $10^{-3}$ cm$^{-3}$ (Landini and Monsignori-Fossi 1990),
the characteristic density of the warm phase in the 1-3' and 3-6' regions for a
volume filling factor $f=1$ ( 0 $<$  $f \leq$ 1). 
Our observations however  determine  the so called {\it emission measure} of the warm phase
($EM \propto \int n^2 dV$, $V$ being the volume of the emitting gas), 
and $f$ remains unknown. 
For a given detected $EM$, the total warm gas mass $M$ in a given region is 
a function of $f$, $M \propto \sqrt{f}$;
 as the density increases with decreasing $f$ ($n \propto 1/ \sqrt{f}$),
the cooling time accordingly decreases as $t_{cool} \propto \sqrt{f}$.
 The mass implications are very demanding: this new component
will have a mass comparable to that of the hot ICM, see Table 2, and those
budgets can
be alleviated if $f$ is very small, however at the price of a correspondingly shorter 
cooling time.

Alternatively, an inverse Compton (IC) origin of the emission can be advocated (Sarazin and Lieu 1998).
Diffusive shock acceleration (Axford et al. 1977; Bell 1978a,b; Blandford 
and Ostriker 1978) may be responsible for a population of
relativistic electrons  with a power-law spectrum in the ICM,  and 
its IC mechanism preserves a similar power-law shape, whereby the two differential
number distributions of electrons (N(E)$\propto E^{-\mu}$) and emitted 
photons (L$(\epsilon) 
\propto \epsilon^{-\alpha}$) are related by the equation $\alpha=(1+\mu)/2$. 
Relativistic electrons of $\gamma$ =300-700, 
required to emit EUV or soft X--ray through IC scattering,
straddle spectrally between low--energy Coulomb losses and high--energy radiative losses 
(Lieu et al. 1999; Sarazin 1999),
and may survive for a significant fraction of a cluster's lifetime.
Here we fit the photon index $\alpha$
to the data, Table 2, with
the exception of the 3-6 arcmin region
 where $\alpha$ it is fixed at $\alpha$=1.75, the expected value from
a cosmic ray (CR) population of $\mu$=2.5 in accordance to
Galactic CR. 
The modelling is formally acceptable, although with a spectral index somewhat
steeper than that of Galactic CR. 
However, serious problems confront  the energetic requirements of the  IC model: the pressure
of CR electrons with $\gamma$ in the range 475--700 (those which emit IC radiation
in the E=0.2-0.4 keV band) exceeds that of the hot ICM by a factor of order 2-3!
Moreover, this pressure estimate is necessarily a lower limit, as 
higher-- and lower--energy electrons may be present, as well as CR ions which, although not 
visible, will further drive the CR budget to absurde values (Lieu et al. 1999; see also Miniati
et al. 2001).

As often happens in rich clusters of galaxies, the X--ray centroid, here coincident with the
central cD galaxy ESO 291-9, is associated with a lower temperature for
the hot ICM (Kaastra et al. 2001). 
Such a temperature decrease is also clearly visible in our PSPC data.
Yet  the detected soft X--ray excess cannot be directly related to it: firstly, because
the excess covers an area about 25 times larger than the region affected by this `cooling' 
(which has a radius of about 1.8 arcmin, Kaastra et al. 2001; Allen and Fabian 1997); 
secondly, and more importantly, 
because we already accounted for the effect in our data modeling (see best--fit
temperatures in Table 1). 
The behaviour here is therefore analogous to, e.g., that of the Virgo and A1795 clusters, 
where the excess emission spreads over a much larger area than that of
the central cooler region (Lieu et al. 1996a; Bonamente et al. 2001a).
The central galaxy of S\'{e}rsic 159-03, which also contains an IR source,
 is well contained within a radius of 1 arcmin (Hansen et al. 2000): the soft excess emission
must then be  a genuinely cluster--wide phenomenon.

\section*{Discussion and conclusions}
The high statistical significance of the soft excess emission of this cluster
confounds both thermal and non-thermal models,
Table 2. Either model can fit  the data,
and calls for an overhaul in 
our understanding of galaxy clusters.

Following the non-thermal interpretation, the large pressure budgets of
Table 2
cannot easily be mitigated; on the contrary, they are strict lower limits
to the relativistic particle content in the cluster.
The presence of a population of relativistic particles
{\it above} equipartition with the hot gas 
 clearly  has implications on a cluster's hydrostatic balance
(e.g., Berezinsky et al. 1997), and the role of intra-cluster
 CR  on the cluster's evolution (e.g., heating of the ICM gas)
 could be much more significant than  previously thought. At present it is not known
whether strong shocks, e.g., those induced by cluster mergers, may accelerate
 such a large amount of CRs. It is possible that the cluster experiences many such
shocks throughout its lifetime; the ICM would then accumulate
those
CR  electrons  with the longest radiative life time, precisely  
the $\gamma \sim$ 300-500 electrons which emit in EUV and soft X-rays through 
IC scattering (Sarazin
and Lieu 1998; Lieu et al. 1999).

On the other hand, large warm gas masses could be sustained
at the interface of cold gas clouds and the hot ICM (Fabian 1997), 
the presence of a `cold' (T $\leq 10^4$ K) phase in the ICM
is in fact an interpretation of the PSPC images of the Coma and Virgo clusters
(Bonamente et al. 2001c). In the absence of  such a replenishment  mechanism, however,
the very short radiative cooling time renders the thermal model untenable.

The bright soft excess of S\'{e}rsic 159-03 here reported
presents a new and unavoidable reality: whether the ultimate explanation be thermal,   
non-thermal or some other origin, a 
major effect at work in the intergalactic medium has hiterto been
completely ignored.

\newpage 

\section*{References}

\noindent 
Abell, G.O., Corwin, H.G. and Olowin, R.P. 1989, {\it ApJS}, {\bf 70}, 1.\\
Allen, S.W. and Fabian, A.C. 1997, {\it MNRAS}, {\bf 286}, 583. \\      
\noindent
Anders, E. and Grevesse, N. 1989, {\it Geochim. Cosmochim. Acta}, {\bf 53}, 197. \\
\noindent Arabadjis J.S. \& Bregman, J.N., 1999, {\it ApJ},
{\bf 514}, 607. \\  
\noindent
Arnaud, M. et al. 2001, {\it A \& A}, {\bf 365}, L80.\\ 
\noindent
Axford, W.I., Leer, E. and Skadron, G. 1977, {\it Proc. 15th Int. Cosmic-Ray \\
\indent Conference (Plovdiv)}, {\bf 11}, 132. \\ 
\noindent
Bell, A.R. 1978a, {\it MNRAS}, {\bf 182}, 147.\\
\noindent
Bell, A.R. 1978b, {\it MNRAS}, {\bf 182}, 443.\\
\noindent
Berezinski, V.S., Blasi, P. and Ptuskin, V.V 1997, {\it ApJ}, {\bf 487}, 529.\\
\noindent 
Berghoefer, T.W., Bowyer, S., \& Korpela, E.J., 2000, {\it ApJ},
{\bf 545}, 695. \\
\noindent 
Berghoefer, T.W., Bowyer, S., \& Korpela, E.J., 2000, {\it ApJ},
{\bf 535}, 615. \\                                                                    
\noindent
Blandford, R.D. and Ostriker, J.P. 1978, \it A \& A \rm, {\bf 221}, L29. \\    
\noindent
Bonamente, M., Lieu, R. and Mittaz, J.P.D. 2001a, \it ApJ \rm, {\bf 547}, L7. \\
\noindent
Bonamente, M., Lieu, R. , Nevalainen, J. and Kaastra, J.S. 2001b, \\
\indent \it ApJ\rm, {\bf 552}, L7. \\          
\noindent
Bonamente, M., Lieu, R. and Mittaz, J.P.D. 2001c, \it ApJ \rm, {\bf 546}, 805.\\
\noindent
Boulanger, F. and Perault, M. 1988, {\it ApJ}, {\bf 330}, 964. \\   
\noindent
Bowyer, S., Berghoeffer, T.W. 1998, {\it ApJ}, {\bf 506}, 502. \\
\noindent
Bowyer, S., Berghoeffer, T.W., \& Korpela, E.J., 1999, {\bf ApJ},
{\it 526}, 592.  \\
\noindent 
Bowyer, S., Korpela, E.J., \& Berghoefer, T.W., 2001, {\it ApJ}, {\bf 548},
L135. \\  
\noindent
Buote, D.A., 2000, {\it ApJ}, {\bf 544}, 242.\\
\noindent 
Buote, D.A., 2000, {\it ApJ}, {\bf 532}, L113.\\
\noindent 
~Buote, D.A., 2001, {\it ApJ}, {\bf 548}, 652. \\                              
\noindent
Chandran, B.D.G., Cowley, S.C and Allbright, B. 1999, {\it Diffuse thermal and \\
\indent relativistic plasma in galaxy clusters}, MPI, Garching, 242. \\    
\noindent
Cowie, L.L. and McKee, C.F. 1977, {\it ApJ}, {\bf 211}, 135. \\ 
\noindent
De Grandi, S. et al. 1999,  {\it ApJ}, {\bf 514}, 148. \\
\noindent
Dickey., J.M. and Lockman, F.J. 1990, {\it Ann. R. Astron. Astrop.},
{\bf 28}, 215. \\
\noindent 
Dixon, W.V., Sallmen, S., Hurwitz, M., \& Lieu, R., 2001,
{\it ApJ}, {\bf 550}, L25.\\
\noindent 
Dixon, W.V., Hurwitz, M., \& Ferguson, H.C., 1996, {\it ApJ},
{\bf 469}, L77. \\   
\noindent Fabian, A.C. 1996, {\it Science}, {\bf 271}, 1244. \\
\noindent Fabian, A.C. 1997, {\it Science}, {\bf 275}, 48. \\   
\noindent 
Fusco-Femiano, R., Dal Fiume, D., et al, 2000, {\it ApJ}, {\bf 534}, L7. \\ 
\noindent
Hansen, H.E., J\o rgensen. H.E., N\o rgaard-Nielsen, H.U, Pedersen, K., \\
\indent Goudfrooij, P and Linden-V\o rnle, M.J.D. 2000, {\it A \& A}, {\bf 362}, 1018.\\ 
\noindent
~Kaastra, J.S. 1992 in \it An X-Ray Spectral Code for Optically Thin Plasmas \rm
 \\\indent
(Internal SRON-Leiden Report, updated version 2.0). \\    
\noindent
Kaastra, J.S., Lieu, R., Mittaz, J.P.D. et al. 1999, {\it ApJ}, {\bf 519}, L119. \\ 
\noindent
Kaastra, J.S., Ferrigno, C., Tanura, T., Paerels, F.B.S., Peterson, J.R. \\
\indent and Mittaz, J.P.D. 2001, \it A \& A \rm, {\bf 365}, 99. \\   
\noindent
Landini, M. and Monsignori-Fossi B.C. 1990, {\it A \& AS}, {\bf 82}, 229.\\
\noindent
Krick, J., Arabadjis, J.N. and Bregman, J.N. 2000, {\it BAAS}, {\bf 197}, 07.11. \\
\noindent
Lieu, R., Mittaz, J.P.D., Bowyer, S., Lockman, F.J.,
Hwang, C. -Y., Schmitt, \\
\indent  J.H.M.M. 1996a, \it ApJ \rm, {\bf 458}, L5. \\
\noindent
Lieu, R., Mittaz, J.P.D., Bowyer, S., Breen, J.O.,
Lockman, F.J., \\
\indent Murphy, E.M. \& Hwang, C. -Y. 1996b, {\it Science}, {\bf 274},1335--1338. \\
\noindent
Lieu, R., Bonamente, M., Mittaz, J.P.D., Durret, F., Dos Santos, S. and \\
\indent Kaastra, J. 1999a, ApJ, {\bf 527}, L77.\\
\noindent
Lieu, R., Bonamente, M. and Mittaz, J.P.D. 1999,
{\it ApJ}, {\bf 517}, L91.\\ 
\noindent Lieu, R., Bonamente, M., \& Mittaz, J.P.D., 2000, {\it A \& A}, {\bf 364},
497.\\   
\noindent
Lieu, R., Ip, W.-I., Axford, W.I. and Bonamente, M. 1999, {\it ApJ}, {\bf 510},
L25.\\
\noindent
~Mewe, R., Gronenschild, E.H.B.M., and van den Oord, G.H.J., 1985 \\
\indent {\it A\&A Supp.}, {\bf 62}, 197.  \\
\noindent Mewe, R., Lemen, J.R., and van den Oord, G.H.J. 1986,
\it A\&A. Supp.\rm, \\
\indent {\bf 65}, 511--536. \\   
\noindent
Miniati, F., Ryu, D., Kang, H. and Jones, T.W. 2001, \it ApJ \rm in press. \\ 
\noindent
Mittaz, J.P.D., Lieu, R. and Lockman, F.J. 1998, \it ApJ\rm, 498, L17. \\
\noindent
Reynolds, A.P et al. 1999,
{\it A \& AS}, {\bf 134}, 287.\\ 
\noindent
 Morrison, R. and McCammon, D. 1983, {\it ApJ}, {\bf 270}, 119.\\    
\noindent
Sarazin, C.L. and Lieu, R. 1998, {\it ApJ}, {\bf 494}, L177. \\                               
\noindent
Sarazin, C.L. 1999, {\it ApJ}, {\bf 520}, 529.\\                  
\noindent
Smith, D.F. and Lilliequist, C.G. 1979, {\it ApJ}, {\bf 232}, 582. \\
\noindent Valinia, A. Arnaud, K., Loewenstein, M., et al. 2000,
{\it ApJ}, {\bf 541}, 550.\\  
\noindent Yan, M., Sadeghpour, H.R. and Dalgarno, A. 1998, {\it ApJ}, {\bf 496}, 1044.\\       
\noindent
Wilms, J., Allen, A. and MCCray R. 2000, {\it ApJ}, {\bf 542}, 914.\\
\noindent
Wheelock, S. et al., 1994, {\it IRAS Sky Survey Expl. Supp.}, (IPL Publ. 94-11), \\
\indent Pasadena, CA. \\
\noindent
Wu, X.-P., Xue, Y.-J. and Fang, L.-Z. 1999,  {\it ApJ}, {\bf 524}, 22. \\

\newpage

\begin{table}[h!]
\begin{center}
\tighten
\caption{Spectral modelling of PSPC spectra of S\'{e}rsic 159-03 with the simple
1-T model. Here and after errors are 90 \% confidence ($\chi^2$ + 2.7 method),
where no errors are reported parameters were held fixed, according to the X--ray study of
Kaastra et al. (2001). Fit is performed by minimizing the $\chi^2$ function, d.o.f. is degrees
of freedom. `A' is elemental abundance following the relative proportions of Anders and Grevesse 
(1989). Photoelectric absorption code is MM83, in square brackets the new WAM00
code is employed, with same results as MM83.
}
\small
\vspace{2cm}
\hspace{-1cm}
\begin{tiny}
\begin{tabular}{lcccccccccc}
\hline
Region & \multicolumn{3}{c}{0.2-2.0 keV fit} &\multicolumn{4}{c}{0.2-2.0 keV fit, free NH}  &
  \multicolumn{3}{c}{0.5-2.0 keV fit}  \\ 
(arcm.)    & k$T$ & A & $\chi^2$(d.o.f) & k$T$ & A & NH & $\chi^2$(d.o.f) & k$T$ & A & $\chi^2$(d.o.f) \\
 & (keV) & & (keV)& & $10^{20}$ cm$^{-2}$ & & & (keV) & & \\
  &  \multicolumn{3}{c}{\hrulefill} & \multicolumn{4}{c}{\hrulefill} & \multicolumn{3}{c}{\hrulefill} \\
0-1 & 1.44 $\pm^{0.1}_{0.02}$ & 0.08 $\pm^{0.02}_{0.02}$ & 264(167) &
      1.88 $\pm^{0.23}_{0.11}$ & 0.4 $\pm^{0.12}_{0.09}$ & 0.95 $\pm^{0.12}_{0.11}$ & 143(166)&
      1.87 $\pm^{0.14}_{0.11}$ & 0.41 $\pm^{0.14}_{0.1}$ & 97(137) \\
 & [ 1.48 $\pm 0.07$ &   0.08 $\pm^{0.015}_{0.02}$& 272(167) & 
       1.92 $\pm^{0.12}_{0.14}$& 0.43 $\pm^{0.14}_{0.1}$ &   0.92 $\pm^{0.12}_{0.13}$ &142(166) & 
       1.87 $\pm^{0.14}_{0.11}$ & 0.41 $\pm^{0.14}_{0.1}$ & 97(137) ] \\ 
1-3 & 1.38 $\pm{0.08}$ & 0.03 $\pm^{0.02}_{0.03}$ & 335(162) &
      2.9$\pm^{0.9}_{0.4}$ & 0.5$\pm^{0.34}_{0.21}$ & 0.52 $\pm^{0.13}_{0.14}$ & 154(161) &
      2.9$\pm^{0.7}_{0.5}$ & 0.5$\pm^{0.37}_{0.23}$ & 115(132) \\
 &  [ 1.38 $\pm^{0.08}_{0.1}$ & 0 $\pm^{0.05}_{0}$ & 342(162) &
     3 $\pm^{0.8}_{0.5}$ & 0.5  $\pm^{0.36}_{0.21}$ & 0.51  $\pm^{0.14}_{0.15}$ & 153(161) &
     2.9  $\pm{0.65}$ & 0.5  $\pm^{0.38}_{0.23}$ & 115(132) ]\\
3-6 & 2.0$\pm^{0.5}_{0.36}$ & 0.02 $\pm^{0.08}_{0.02}$ & 111(128) &
      4.0$\pm^{3.8}_{1.4}$ & 0.6$\pm^{6}_{0.5}$ & 0.7$\pm^{0.3}_{0.5}$ & 88(127) &
      3.8$\pm^{4}_{1.4}$ & 0.6$\pm^{5}_{0.5}$ & 66(99) \\
 & [ 2 $\pm^{0.45}_{0.76}$ & 0 $\pm^{0.01}_{0}$ & 112(128)&
     4.1 $\pm^{3.8}_{1.4}$ & 0.77 $\pm^{5}_{0.6}$ & 0.68 $\pm^{0.32}_{0.45}$& 88(127) &
    3.7 $\pm^{1.9}_{1.2}$ & 0.64 $\pm^{5}_{0.63}$ & 66(99) ] \\
6-9 & 2 & 0.3 & 67(84) & 2 & 0.3 & 1.3$\pm^{0.8}_{0.65}$ & 65(83) & 2 & 0.3 & 31(55) \\  
    &[ 2 & 0.3& 67(84) & 2 & 0.3 & 1.3 $\pm^{0.85}_{0.4}$ & 66(83) & 2 & 0.3 & 31(55) ] \\
\hline
\end{tabular}
\end{tiny}
\end{center}
\end{table}

\newpage 

\begin{table}[h!]
\begin{center}
\tighten
\caption{Spectral modelling with a 2-T model (`hot' and `warm' phases of ICM) and with a
non--thermal model. Mass implications for the former and pressure budgets for the latter
are reported; $f$ is filling factor of the warm component. See Table 1 for explanation of
Galactic absorption codes employed.} 
\vspace{2cm}
\hspace{-1cm}
\begin{tiny}
\begin{tabular}{lccccccccccc}
\hline
Region & \multicolumn{5}{c}{2-T model} & \multicolumn{6}{c}{Non-thermal model} \\
(arcm.) & k$T_h$ & A & k$T_w$ & $\chi^2$(d.o.f) & $M_w/M_h$ & k$T$ & A & $\alpha $ & $\chi^2$(d.o.f) &
$P_{therm}$ & $P_{IC}$ \\
 & (keV) & & (keV) & & &(keV) & & & & \multicolumn{2}{c}{$10^{-11}$ (erg cm$^{-3})$} \\
    & \multicolumn{5}{c}{\hrulefill} & \multicolumn{6}{c}{\hrulefill} \\
0-1 & 1.88$\pm^{0.33}_{0.11}$ & 0.43$\pm^{0.32}_{0.07}$ & 0.078$\pm^{0.035}_{0.015}$ & 130(165) &
  $\sqrt{f} \times$ 0.38$\pm^{0.25}_{0.05}$ & 1.94$\pm^{0.16}_{0.13}$ & 0.5 & 3.2$\pm^{0.9}_{0.2}$ &
  136(166) & 7 & 22$\pm^{25}_{13}$ \\
  & [ 1.9 $\pm^{0.2}_{0.12}$ & 0.64$\pm^{00.15}_{0.11}$ & 0.07$\pm{0.035}$& 129(165)&
  $\sqrt{f} \times$ 0.34 $\pm^{0.26}_{0.17}$ & 1.94$\pm^{0.16}_{0.13}$ & 0.5 & 3.2$\pm^{1.1}_{0.3}$ &
  135(166) & & 20 $\pm^{22}_{12}$ ] \\
1-3 & 2.9 $\pm^{0.74}_{0.46}$ & 0.53$\pm^{0.47}_{0.21}$ & 0.06$\pm^{0.04}_{0.021}$& 142(160) &
  $\sqrt{f} \times$ 0.55$\pm^{0.6}_{0.2}$ & 2.8$\pm^{0.8}_{0.4}$ & 0.4 & 4.8$\pm^{1.2}_{0.6}$ & 
  148(161) & 0.7 & 2.3 $ \pm^{7.5}_{1.5} $ \\
 &[ 3$\pm^{0.85}_{0.5}$ & 0.67 $\pm^{0.68}_{0.32}$ & 0.07 $\pm^{0.035}_{0.02}$ & 141(160) &
  $\sqrt{f} \times$ 0.33 $\pm^{0.82}_{0.22}$ & 2.8$\pm^{0.8}_{0.4}$ & 0.4 & 4.7 $\pm^{1.2}_{0.2}$ &
  148(161) & & 2 $\pm^{7}_{1.3}$ ] \\
3-6 & 3.5$\pm^{2.7}_{1.2}$ & 0.3 & 0.06$\pm{0.0035}$ & 88(127) &
  $\sqrt{f} \times$ 0.46$\pm^{15}_{0.26}$ & 2.5 & 0.2 & 1.75 & 
  96(129) & 0.14 & 0.33 $\pm^{0.1}_{0.11}$ \\ 
  & [ 3.5 $\pm^{3.1}_{1.2}$ & 0.3 & 0.055$\pm^{0.06}_{0.03}$ & 88(127) &
  $\sqrt{f} \times$ 0.61 $\pm^{17}_{0.42}$ & 2.5 & 0.2 & 1.75 & 
  96(129) & & 0.36 $\pm{0.12}$ ] \\ 
\hline
\end{tabular}
\end{tiny}
\end{center}
\end{table}

\newpage

Figure 1: Radial profiles of X-ray brightness in PSPC R2 (0.2-0.4 keV, red) and R37 
(0.4-2.0 keV, green)
bands in units of counts/s/sq.arcmin, the lines mark the corresponding background levels.
Background is determined from an annulus $\sim$ 40' off-axis.

Figure 2: Three PSPC spectra of the central region (radius 3 arcmin) of S\'{e}rsic 159-03. The data,
available through the public archive (identification numbers 800397, 600182 and 600429a1), are
individually rebinned to obtain a minimum S/N $>$ 4 in each spectral channel. The solid lines describe the 
best--fit single temperature model (see text for description) that highlights the low--energy
excess of photons. Agreement among the 3 observations is excellent, see bottom panel. 

Figure 3: Coadded spectrum of 1-3 arcmin region of the cluster (black crosses) overlaid
on the best fit 1-T model obtained by fitting the whole PSPC band (left, red solid line) 
and by fitting only energies $\geq$ 0.5 keV (right, green solid line). Residuals (red and green crosses)
reveal in both cases excess of soft photons at energies $<$ 0.4 keV.

Figure 4: Fractional excess in 1/4 keV  PSPC band (0.2-0.4 keV) as function of radial
distance from the cluster's center. Fractional excess is defined as $\eta=(m-p)/p$, where $m$ is measured
soft flux (1/4 keV band) and $p$ its prediction according to the 0.5-2.0 keV single temperature model.
The 9-12' region was modelled with the same parameters of the 6-9' annulus (Table 1);
vertical semi-diameters are 1-$\sigma$ statistical errors.
The excess component ranges between 50 and 100 \% of the hot ICM gas in the soft band, the
strongest observed to date in PSPC data.


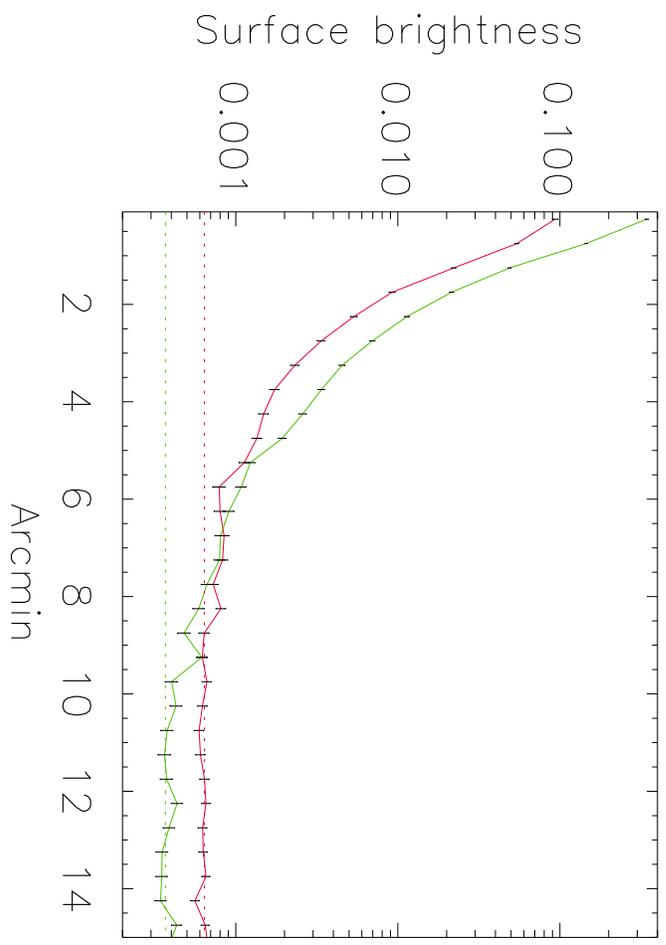
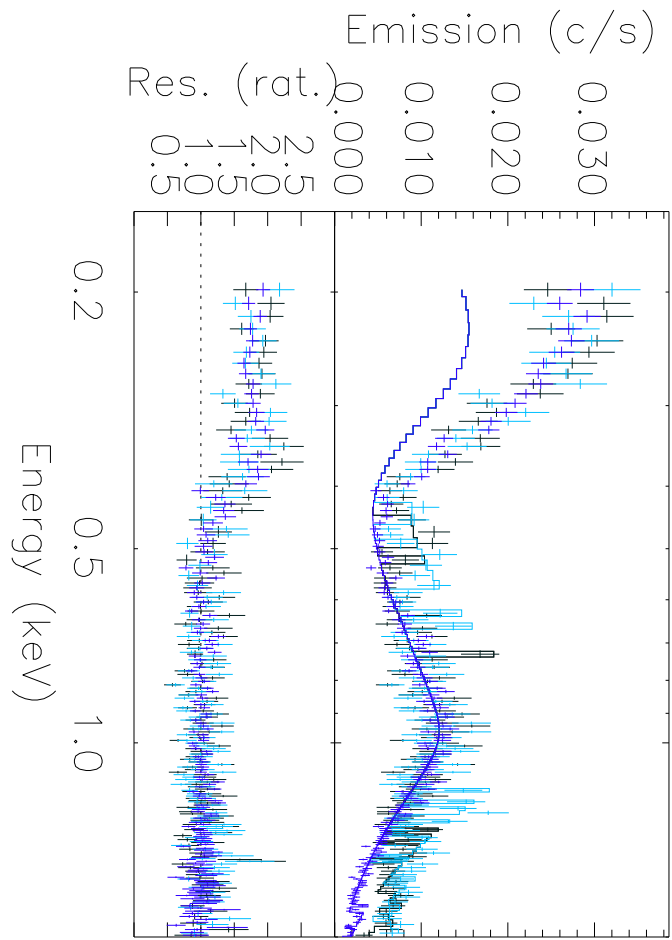
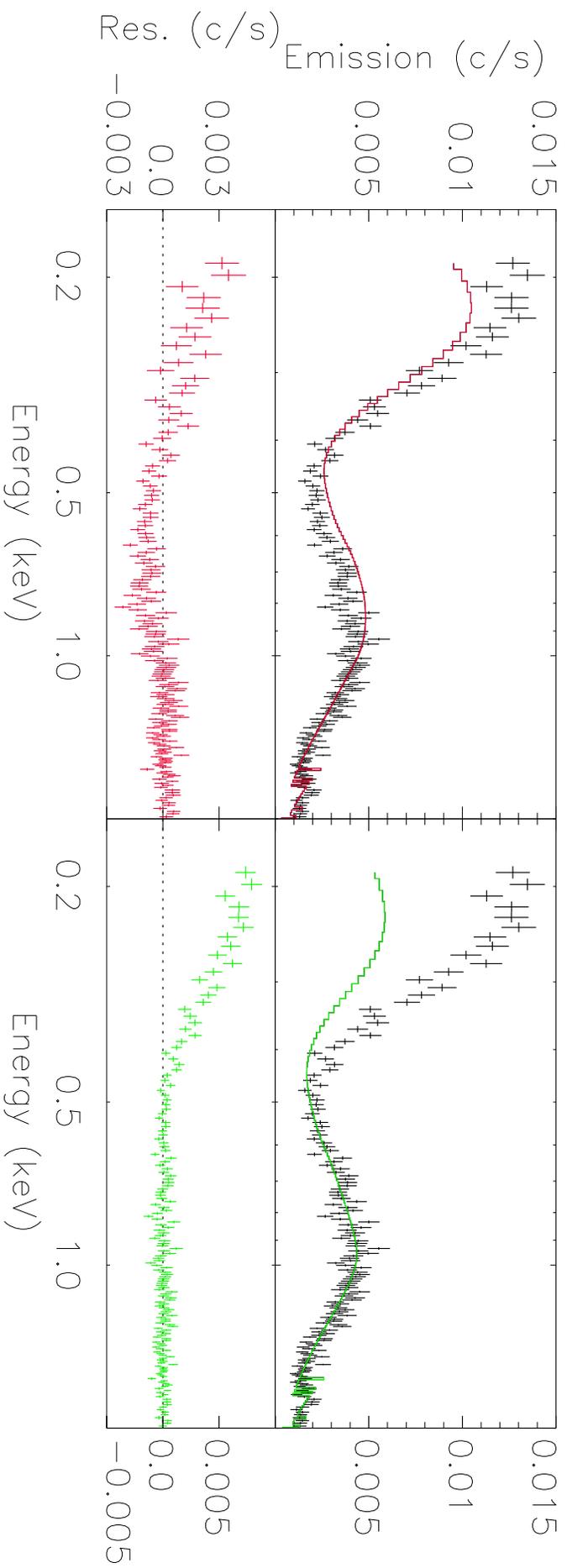
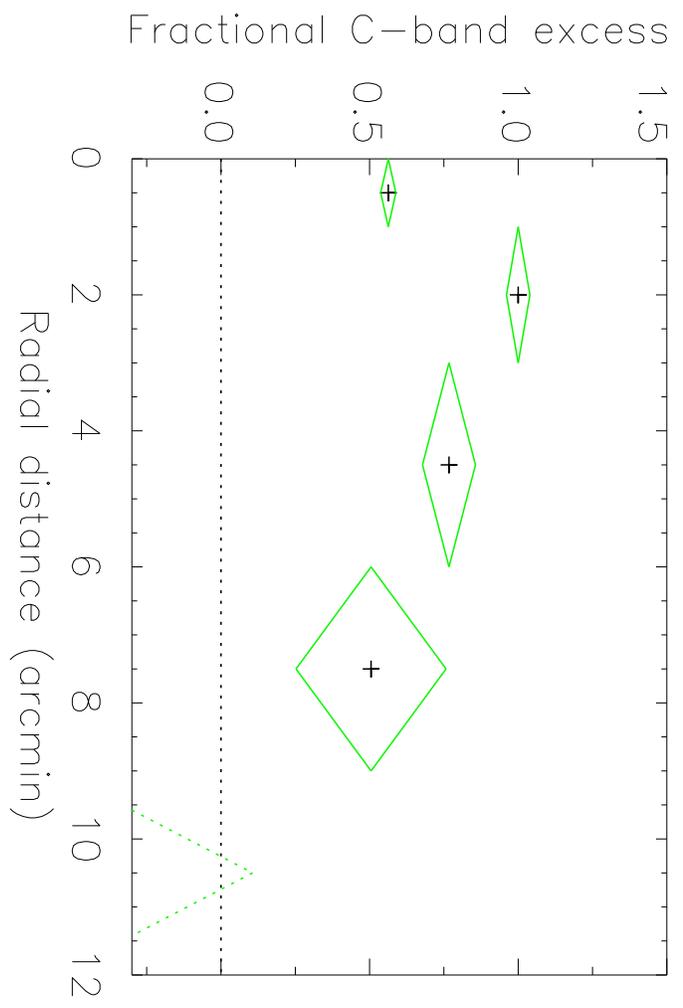
\end{document}